# Old Drugs for Newly Emerging Viral Disease, COVID-19: Bioinformatic Prospective

**Mohammad Reza Dayer**

**Department of Biology, Faculty of Science, Shahid Chamran University of Ahvaz, Ahvaz, Iran**

**Corresponding author:**

**Mohammad Reza Dayer, Department of Biology, Faculty of Sciences, Shahid Chamran University of Ahvaz, Ahvaz, Iran. Tel/Fax: +98-6113331045, E-mail: mrdayer@scu.ac.ir**

**Abstract:**

Coronavirus (COVID-19) outbreak in late 2019 and 2020 comprises a serious and more likely a pandemic threat worldwide. Given that the disease has not approved vaccines or drugs up to now, any efforts for drug design and or clinical trails of old drugs based on their mechanism of action are worthy and creditable in such circumstances. Experienced docking experiments using the newly released coordinate structure for COVID-19 protease as a receptor and thoughtfully selected chemicals among antiviral and antibiotics drugs as ligands may be leading in this context. We selected nine drugs from HIV-1 protease inhibitors and twenty-one candidates from anti bronchitis drugs based on their chemical structures and enrolled them in blind and active site-directed dockings in different modes and in native-like conditions of interactions. Our findings suggest the binding capacity and the inhibitory potency of candidates are as follows Tipranavir >Indinavir>Atazanavir>Darunavir>Ritonavir>Amprenavir for HIV-1 protease inhibitors and Cefditoren>Cefixime>Erythromycin>Clarithromycin for anti bronchitis medicines. The drugs bioavailability, their hydrophobicity and the hydrophobic properties of their binding sites and also the rates of their metabolisms and deactivations in the human body are the next determinants for their overall effects on viral infections, the net results that should survey by clinical trials to assess their therapeutic usefulness for coronavirus infections.

## Introduction

The outbreak of coronavirus (COVID-19) in 2019-2020 is caused by Severe Acute Respiratory Syndrome Corona Virus (SARS-COV-2) [1-4]. It is a positive-sense single-stranded RNA virus that caused a total of 75,465 reported cases up to February 2020 in China [5-6]. Increased risk of world wide spread comprises serious life-threatening issues against human safety. Fever, cough, and shortness of breath are the main symptoms of the disease that may leads eventually to pneumonia with a mortality rate of 1-3% [7-11]. Currently, there are no approved drugs for coronavirus infectious. Even though antiviral drugs such as inhibitors against protease, integrase and or polymerase enzymes designed and are in advance studies [12-13]. Among these inhibitors, anti-protease inhibitors seem to act effectively in blocking virus replication and provide a promising treatment for SARS and MERS diseases.

Given the protease vital role in the virus life cycle and its maturation via functional protein production from their precursor, it seems to be a good target for drug design in viral infections as COVID-19 infection as well. Based on data available in the PUBMED database (https://www.ncbi.nlm.nih.gov/pubmed/) COVID-19 protease, EC 3.4.2 is a protein with 305 residues contrast to SARS protease with 306 amino acid residues. Sequence alignment using EMBOSS Stretcher server (www.ebi.ac.uk) reviled that corona protease compared to SARS protease contains one amino acid deletion at position 144 (Cys) as well as about 10 mutations along it sequence stretch as depicted in scheme 1. These changes in COVID-19 protease sequence may lead to the different global architecture of protein and especially in its binding or catalytic site, in such a way the localization of binding sites in COVID-19 shifted to neighbors' residues contrast to that's of SARS protease.

Binding site survey for these two proteases using Computed Atlas of Surface Topography of proteins (http://sts.bioe.uic.edu/castp/) server confirmed these changes in enzyme binding site. Table 1 represents the binding site residues for SARS and COVID-19 protease. As it is evident there is only three residues similarity between two enzymes at positions 140-142 different amino acid constituent and geometry expectedly need different inhibitors with different stereochemistry. Based on this fact it is necessary to search for different inhibitors for coronavirus, the investigation that is the main topic of this study.

**Table 1: Active site residues extracted from Computed Atlas of Surface Topography of proteins (http://sts.bioe.uic.edu/castp/) server for COVID-19 and SARS proteases.**

| **SARS** | **SARS(Continued)** | **COVID-19-Crystal Structure** | **COVID-19-Optimised Structure** |
|---|---|---|---|
| Residue(No) | Residue(No) | Residue(No) | Residue(No) |
| PHE(3) | **ASN(142)** | THR(24) | |
| ARG(4) | ILE(152) | **THR(25)** | **THR(25)** |
| LYS(5) | ASP(153) | THR(26) | |
| MET(6) | TYR(154) | **LEU(27)** | **LEU(27)** |
| ALA(7) | ASP(155) | PRO(39) | |
| PHE(8) | GLU(290) | **HIS(41)** | **HIS(41)** |
| PRO(9) | PHE(291) | **CYS(44)** | **CYS(44)** |
| GLY(11) | PHE(294) | THR(45) | |
| LYS(12) | ASP(295) | SER(46) | |
| GLU(14) | VAL(297) | **MET(49)** | **MET(49)** |
| GLY(15) | ARG(298) | **PRO(52)** | **PRO(52)** |
| CYS(16) | GLN(299) | **TYR(54)** | **TYR(54)** |
| MET(17) | CYS(300) | **PHE(140)** | **PHE(140)** |
| VAL(18) | SER(301) | **LEU(141)** | **LEU(141)** |
| TRP(31) | GLY(302) | **ASN(142)** | **ASN(142)** |
| ALA(70) | VAL(303) | **GLY(143)** | **GLY(143)** |
| GLY(71) | | **SER(144)** | **SER(144)** |
| ASN(72) | | | GLY(146) |
| ASN(95) | | **HIS(163)** | **HIS(163)** |
| LYS(97) | | HIS(164) | |
| PRO(99) | | **MET(165)** | **MET(165)** |
| ALA(116) | | **GLU(166)** | **GLU(166)** |
| TYR(118) | | LEU(167) | |
| GLY(120) | | **PRO(168)** | **PRO(168)** |
| SER(121) | | HIS(172) | |
| PRO(122) | | **ASP(187)** | **ASP(187)** |
| SER(123) | | **ARG(188)** | **ARG(188)** |
| GLY(124) | | **GLN(189)** | **GLN(189)** |
| SER(139) | | **THR(190)** | **THR(190)** |
| **PHE(140)** | | | ALA(191) |
| **LEU(141)** | | **GLN(192)** | **GLN(192)** |

**Method and Materials:**

Coordinate structures retrieval and preparations: Coordinate structures of coronavirus protease with PDB ID 6LU7 in accordance with SARS protease with PDB ID 1UK3 were retrieved from protein data bank (https://www.rcsb.org/). The structures, which were obtained by the X-ray diffraction and refined at the resolutions of 2.16 Å and 2.4 Å respectively. In this process, the constructed structure was energy-minimized.3.4. The structures were placed in separate rectangular boxer with dimensions of 8.15×9.06×9.58 nm and 9.44×9.26×10.63nm dimensions respectively. The two boxes were then filled with SPCE solvents with a water shell of 1.0-nm thickness. Steepest descent algorithm was used to minimize the system energy to lower than 200 kJ/mol. Energy minimization was performed at neutral pH (Asp, Glu, Arg and Lys ionized), 37°C and one atmospheric pressure [14-15].

Sequence Alignment was carried out for the two sequences of COVID-19 and SARS protease on EMBOSS Stretcher (www.ebi.ac.uk) for comparison purposes [16-17].

Coordinate structures of enrolled drugs including HIV-1 protease inhibitors of Amprenavir, Atazanavir, Darunavir, Indinavir, Lopinavir, Nelfinavir, Ritonavir, Saquinavir and Tipranavir as well as drugs used to treat bronchitis as Azithromycin, Cefaclor, Cefdinir, Cefditoren, Cefpodoxime, Cefprozil, Ceftriaxone, Cefuroxime, Cifexim, Ciprofloxacin, Clarithromycin, Co-Amoxiclav components (Amoxicillin and Clavulanic acid), Co-Trimoxazole components (Trimethoprim and Sulfamethoxazole), Dicloxacillin, Doxyxycline, Erythromycin, Gemifloxacin, Guaigensin, Moxifloxacin, Ofloxacin, and Tetracycline were obtained from PubChem database (https://pubchem.ncbi.nlm.nih.gov/) in SDF format and converted to PDB format using Open Babel server (http://openbabel.org/). The structures then transferred to ArgusLab software (http://www.arguslab.com/) [18] and checked for their bonds and energy optimization and finally the structures were opened in a text editor in Linux environment for their consistency from right character used point of view.

**Blind Docking experiments:** to survey the potential anchoring site on COVID-19 protease for drugs binding and to verify them as binding sites based on their similarity and vicinity to enzyme active site we performed blind docking experiments in Hex 8.0.0 (http://www.loria.fr/~ritchied/hex/) [19] installed in Linux operating system. The default setting for Shape only, Sahpe+Electrostatic and Shape+Electrostatic+DARS with macro sampling were used in separate experiments on optimized structures of protease and drugs as ligands. The best pose and the binding energies of the 100 poses were recorded for statistical analysis.

**Active Site-Directed Docking:** this kind of docking was performed using ArguLab [18] in default setting using Lamarckian Genetic Algorithm with Max Generations: 10000 and binding site size: 17.96×19.78×26.44 angstroms. The binding energies for the best 20 pose were extracted in Kcal/Mol for statistical analysis.

**Hydrophobicity of Enzyme Active Site and Drugs:** Hydrophobicity of active site as well as drugs could be considered as a determinant factor for effective interactions between enzyme active site and ligands (drugs) that is an indicator for the extent of enzyme inhibition by ligands. In this study, we calculate the hydrophobic index for enzyme active site residues based on Kyte & Doolittle method and report them as hydrophobic index per residue for reasonable comparison [20]. The Hydrophobicity indexes for drugs were calculated as LogP on the Virtual Computational Chemistry Laboratory server (http://www.vcclab.org/) [21].

**Data Handling and Analysis:** all the numerical data were exploited in Excel and SPSS software. P-value under .05 was considered as the significance level.

**Results and Discussion:**

Sequence alignment for COVID-19 and SARS proteases, scheme 1, performed on EMBOSS Stretcher (www.ebi.ac.uk) reviled a single deletion at position 144 (Cys) in COVID-19 protease contrast to SARS ones and also there are twelve mutations along COVID-19 protease along it sequence stretch as depicted in scheme 1 shown by single our double dots instead of vertical line.

```
SARS       1 SGFRKMAFPSGKVEGCMVQVTCGTTTLNGLWLDDTVYCPRHVICTAEDML     50
             ||||||||||||||||||||||||||||||||||.||||||||||:||||
COVID-19   1 SGFRKMAFPSGKVEGCMVQVTCGTTTLNGLWLDDVVYCPRHVICTSEDML     50

SARS      51 NPNYEDLLIRKSNHSFLVQAGNVQLRVIGHSMQNCLLRLKVDTSNPKTPK    100
             ||||||||||||||:||||||||||||||||||||:|:|||||:||||||
COVID-19  51 NPNYEDLLIRKSNHNFLVQAGNVQLRVIGHSMQNCVLKLKVDTANPKTPK    100

SARS     101 YKFVRIQPGQTFSVLACYNGSPSGVYQCAMRPNHTIKGSFLNGSCGSVGF    150
             |||||||||||||||||||||||||||||||||.|||||||||| |||||
COVID-19 101 YKFVRIQPGQTFSVLACYNGSPSGVYQCAMRPNFTIKGSFLNGS-GSVGF    149

SARS     151 NIDYDCVSFCYMHHMELPTGVHAGTDLEGKFYGPFVDRQTAQAAGTDTTI    200
             |||||||||||||||||||||||||||||.||||||||||||||||||||
COVID-19 150 NIDYDCVSFCYMHHMELPTGVHAGTDLEGNFYGPFVDRQTAQAAGTDTTI    199

SARS     201 TLNVLAWLYAAVINGDRWFLNRFTTTLNDFNLVAMKYNYEPLTQDHVDIL    250
             |:||||||||||||||||||||||||||||||||||||||||||||||||
COVID-19 200 TVNVLAWLYAAVINGDRWFLNRFTTTLNDFNLVAMKYNYEPLTQDHVDIL    249

SARS     251 GPLSAQTGIAVLDMCAALKELLQNGMNGRTILGSTILEDEFTPFDVVRQC    300
             ||||||||||||||||:|||||||||||||||||.:||||||||||||||
COVID-19 250 GPLSAQTGIAVLDMCASLKELLQNGMNGRTILGSALLEDEFTPFDVVRQC    299

SARS     301 SGVTFQ    306
             ||||||
COVID-19 300 SGVTFQ    305
```

**Scheme 1: sequence alignment preformed on EMBOSS Stretcher (www.ebi.ac.uk) server. Corona sequence is highlighted in yellow color**

Accordingly, the mutations in COVID-19 expectedly exert structural changes that may affect enzyme binding site position and functional group constituents. As shown in table 1 the mutations move the biding site to preceding residues in COVID-19 contrast to SARS protease, far from the binding site of SARS and hence the inhibitors for SARS should not affect on COVID-19 protease. Given the fact that active site residues placed at flexible or hot points of the protein chain, we extracted the root mean square fluctuation (RMSF) plot of COBID-19 and SARS protease (Figure 1) molecular dynamic simulation trajectory and plot it as RMSD in nanometer per residues numbers. Residues with higher RMSF comprise the hot or flexible points. As it is shown most residues of COVID-19 active site (table 1) placed predominantly in the 140-192 region, while that's of SARS localized in 290-303 regions. This finding reconfirms our claim regarding the COVID-19 active site.

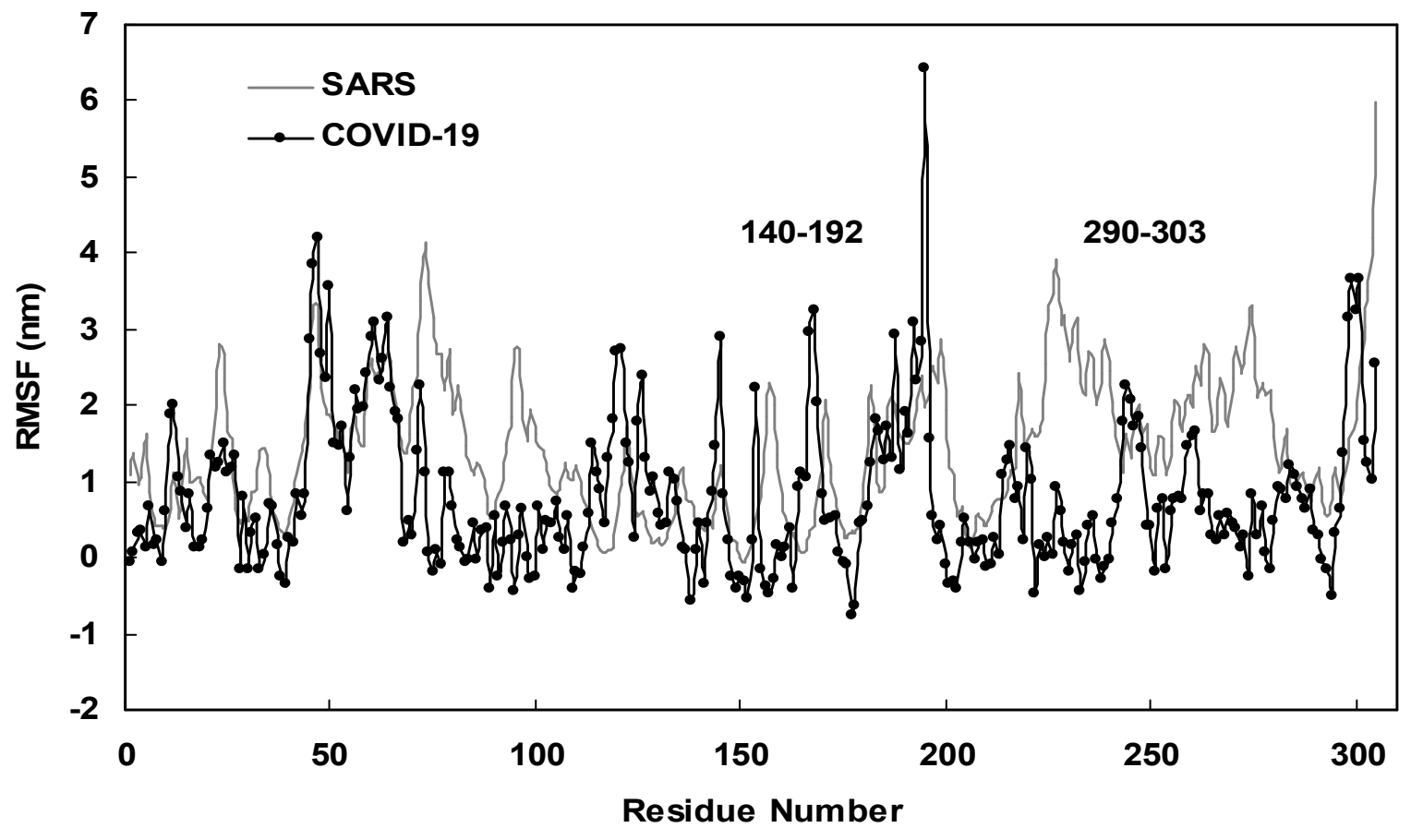

**Figure 1: Root Mean Square Fluctuation of proteases of SARS and COVID-19 obtained from 50ns simulation at 37 degree centigrade, 1 atmosphere of pressure and pH7**

Figure 2 graphically represents the binding patterns of anti-HIV-1 protease (A) include Amprenavir, Atazanavir, Darunavir, Indinavir, Nelfinavir, Ritonavir, Tipranavir, and anti bronchitis drugs (B) Cefditoren, Cifexim, Clarithromycin, Dicloxacillin, Doxycycline, Erythromycin, Tetracycline, Moxifloxacin. As it is indicated all the selected ligands bind to the correct binding site extracted from blind docking experiments with Hex software. This finding reveals that the selected drugs are correct candidates for enzyme inhibition and further study. Lopinavir, Saquinavir against HIV-1 inhibitors did not bind to enzyme active site and among anti bronchitis drugs including Azithromycin, Cefaclor, Cefdinir, Ciprofloxacin, Cefpodoxime, Cefprozil, Ceftriaxone, Cefuroxime, Co-Amoxiclav components (Amoxicillin and Clavulanic acid), Co-Trimoxazole components (Trimethoprim and Sulfamethoxazole), Gemifloxacin, Guaigensin, and Ofloxacin are also did not bind to enzyme active site and accordingly removed from further study.

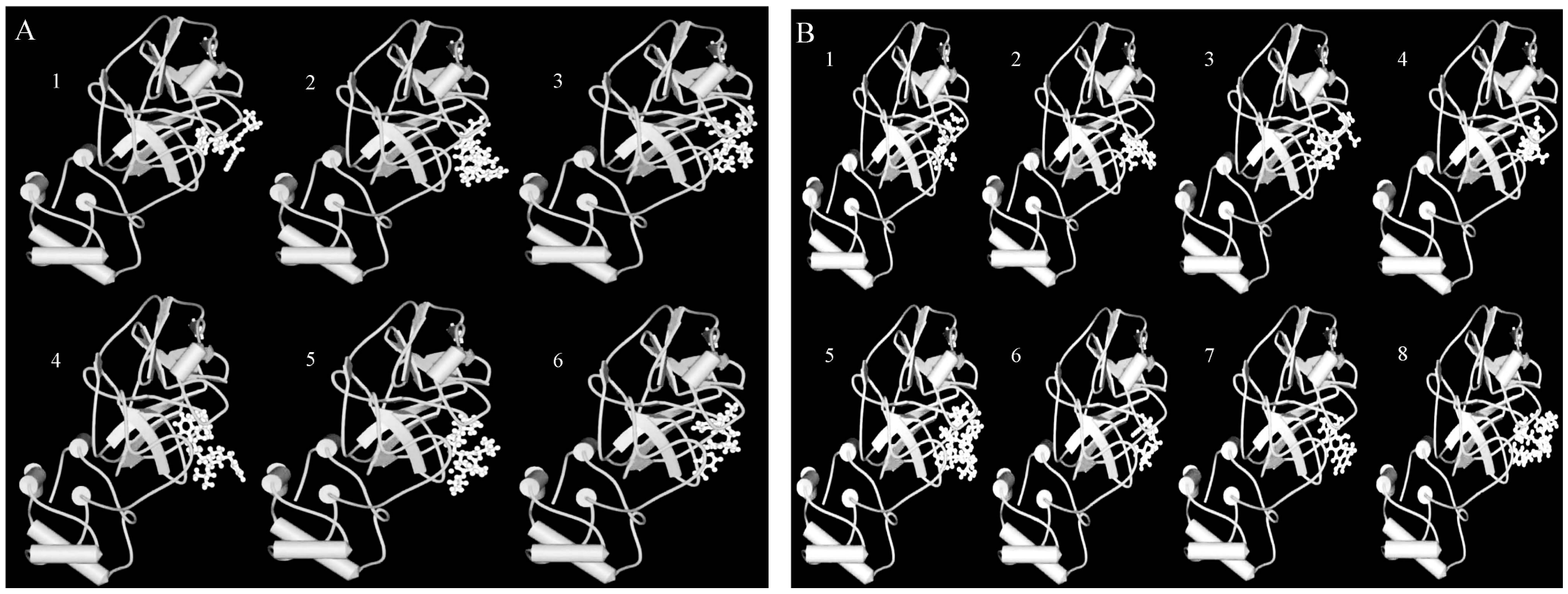

**Figure 2:** Graphically represents the binding patterns of anti HIV-1 protease (A-left) and anti bronchitis drugs (B-right)

In the next step, we carried out active site-directed docking of the selected ligands in ArgusLab software to study the binding capacity of ligands to fit the active site. The binding energies obtained from our docking our blind and active site-directed experiments are shown in table 2. Pearson's two-tailed test of correlation was performed between blind and active site-directed docking reveals that there is no significant correlation between blind and site-directed energies (p-value >.05) and this means that the mechanisms underlie the interactions between ligand and protease is not the same and these two energies are completely independent variables and could be sums for comparative purpose. **Table 2: Binding energies obtained for blind and active site directed docking for studied ligands as well as their total amount (in KJ/Mol) to COVID-19 as receptor**

| | Blind Docking | Active Site | Total Binding Energy |
|---|---|---|---|
| Amprenavir | -227.35 | -29.92 | -257.28 |
| Atazanavir | -253.92 | -27.46 | -281.38 |
| Cefditoren | -258.40 | -32.35 | -290.76 |
| Cefixime | -247.34 | -29.42 | -276.77 |
| Clarithromycin | -265.37 | -6.39 | -271.76 |
| Darunavir | -247.19 | -28.17 | -275.37 |
| Dicloxacillin | -217.90 | -26.96 | -244.86 |
| Doxycycline | -177.37 | -29.76 | -207.14 |
| Erythromycine | -246.92 | -28.08 | -275.01 |
| Ritonavir | -243.61 | -31.60 | -275.21 |
| Indinavir | -249.69 | -35.02 | -284.72 |
| Tetracycline | -216.00 | -26.50 | -242.50 |
| Tipranavir | -269.23 | -40.08 | -309.31 |
| Moxifloxacin | -221.74 | -22.48 | -227.12 |

Figure 3 shows the total binding energies for ligands during our experiments. Based on our findings and as it is clear Tipranavir seems to be the most potent inhibitor for COVID-19 protease. Clinical trials reveal that Tipranavir is capable of viruses' replication in patients resistant to other protease inhibitors and resistance against this drug needs multiple mutations to take place simultaneously in protease genes [22]. The application of the drug in combination with one of the antiretroviral drugs as Ritonavir is approved for HIV-1 treatment by the FDA [23]. The second effective drug we introduced is Cefditoren, a cephalosporin antibiotic with broad-spectrum, administered for different conditions as pneumonia, acute bacterial and chronic bronchitis [24-25]. Indinavir is the third effective inhibitor we candidate for COVID-19 treatment. Considering the miscellaneous drugs available for HIV-1 treatment it is not currently recommended for HIV-1 treatment due to its side effects [26]. However, it is shown that Indinavir by decreasing viral replication increases the life expectancy of the patient for several years in viral infections [27]. The low water solubility of Indinavir requires plenty of water to dink to prevent its precipitation and crystallization as in kidneys [28].

Atazanavir is the forth protease inhibitor that we recommend for COVID-19 infections. This inhibitor structural mimic the transition state of proteolysis at the Phe-Pro cleavage site. Atazanavir is prescribed for HIV-1 treatment in conjunction with an antiretroviral drug [29]. Given that human proteases could not hydrolyze the Phe-Pro bonds so Atazanavir does not inhibit human proteases at all [30-31]. The fifth drug we encourage the clinician's for clinical trials especially that this cephalosporin safely and effectively can combat the probable bacterial infections that may accompany viral infections at facilitates pneumonia and other complications [32-36]

Darunavir is the seventh drug we introduced as COVID-19 protease inhibitors. To increase Darunavir efficacy, it is used in accordance with one of the other retroviral drugs as Reitonavir or cobicistat to treat HIV-1 infectious [37-38]. The drug binds strongly to enzyme active site via multiple hydrogen with a dissociation constant (Kd) of $4.5 \times 10^{-12}$ M, which is much stronger than that's of other protease inhibitors [39]. It is very important to notice that the hydrogen bonds are formed between Darunavir and enzyme backbone and so they did not affected by mutations in protease sequence because the backbone does not affected by sequence and then the inhibitory potency of Darunavir does not disrupted by viral mutations in new generations of viruses [39-40]. The eighth drug we recommend for the clinical trial is Ritonavir which is a known inhibitor for HIV-1 protease. This inhibitor does not or rarely used alone for HIV-1 infectious and instead it is most often used in conjunction with other protease inhibitors to boost their antiviral action. Ritonavir itself inhibits liver and intestine enzymes that metabolize protease inhibitors and by the way, allows lower doses for these drugs to exert their effects [41-44]. The ninth drug seems useful for the clinical trial is Erythromycin the well known and safe antibiotic for respiratory tract infections and finally, Amprenavir is the tenth drug we recommend for a clinical trial for COVID-19 infections that have no vaccine and no approved drug with a reasonable rate of incidence and increasing percent of mortality [45-46].

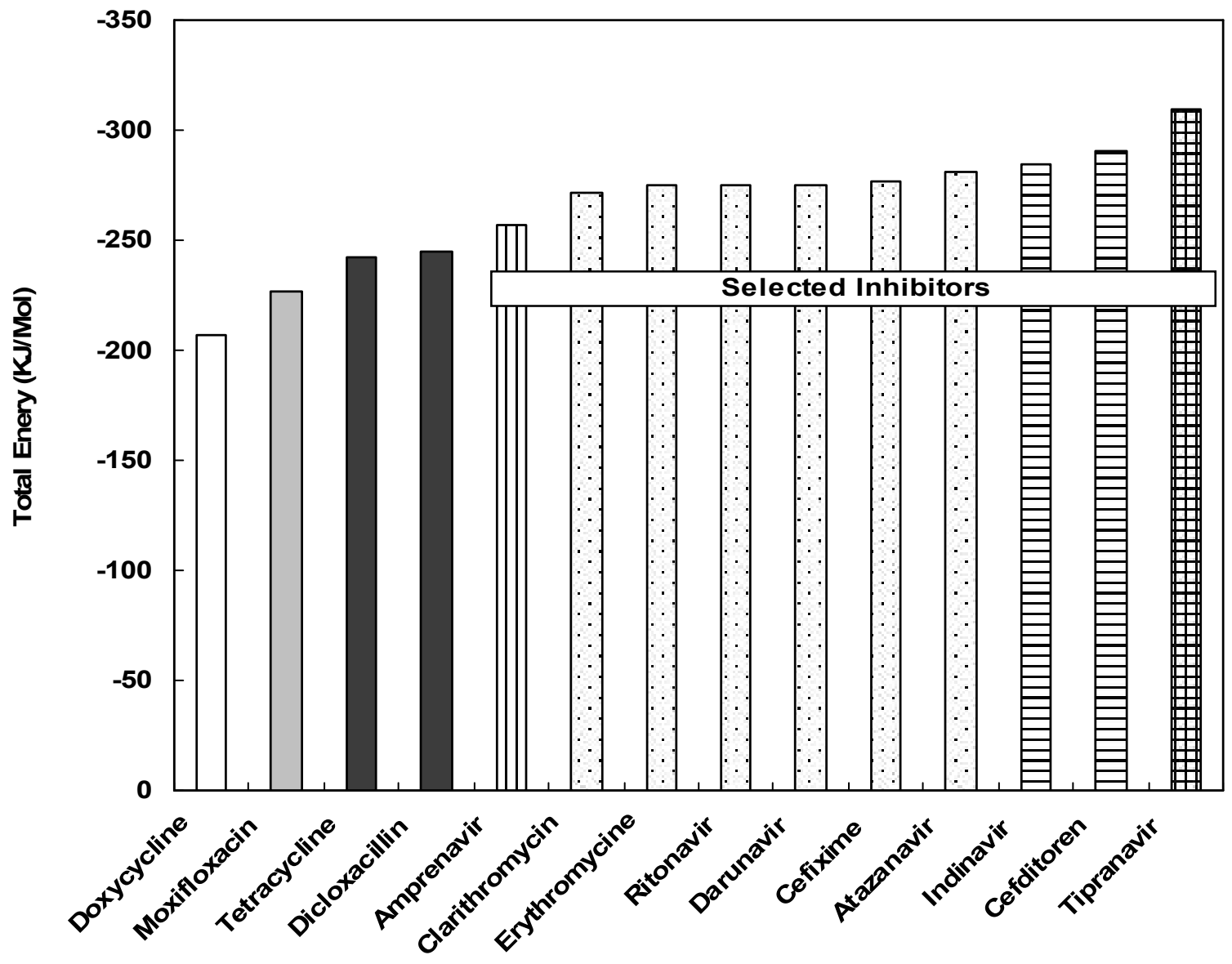

**Figure 3: Total binding energy of drugs in KJ/Mol for drugs. The drugs with not different binding energy are shown with same pattern (p-value>)**

Table 3 represents the complementary parameters that may affect drugs binding to the enzyme active site and their inhibitory effects on enzyme activity. The more similarity in binding site to the enzyme active site cavity causes the more inhibitory effects of drugs on enzyme activity. Among the drugs, Amprenavir exhibits maximum similarity while Darunavir, Dicloxacillin, and Doxycycline show the minimum similarity in their binding site constituents with the enzyme active site. We hypothesize that the binding site similarity for drugs should play determinant roles in their inhibitory potency. Hydrophobicity of drugs binding sites also accelerates their binding in aqueous conditions and we think that the more hydrophobic index for each drug guarantees prominent inhibitory behavior. Atazanavir and Amprenavir based on these criteria should exhibit the most and the less inhibitory character. Finally, LogP calculation for drugs reveals that Tipranavir, Atazanavir, and Ritonavir among anti-HIV-1 drugs and Dicloxacillin, Clarithromycin and Erythromycin among bronchitis drugs manifest the highest hydrophobic properties.

**Table 3: binding site similarity for ligands contrast to enzyme active site and their hydrophobic index per residue in accordance with ligands hydrophobic character of LogP**

| | % of similarity | Hydrophobic Index/Residue | LogP |
|---|---|---|---|
| Amprenavir | 70 | -1.53 | 2.03 |
| Atazanavir | 40 | -0.29 | 4.08 |
| Cefditoren | 50 | -1.70 | 1.7 |
| Cefixime | 65 | -0.79 | 0.25 |
| Clarithromycin | 55 | -1.12 | 3.18 |
| Darunavir | 45 | -1.45 | 1.89 |
| Dicloxacillin | 45 | -0.94 | 3.19 |
| Doxycycline | 45 | -1.28 | -0.98 |
| Erythromycine | 55 | -0.74 | 2.37 |
| Ritonavir | 65 | -0.99 | 4.24 |
| Indinavir | 60 | -1.01 | 3.26 |
| Tetracycline | 60 | -0.83 | -0.56 |
| Tipranavir | 65 | -0.90 | 6.29 |
| Moxifloxacin | 55 | -1.24 | 0.64 |

## Conclusion:

Our results indicate that there are structural differences between primary structures of proteases for COVID-19 contrasts to that's of SARS which leads to altered enzyme binding sites. We hypothesize that these alterations cause the enzyme, not to respond to anti-SARS protease inhibitors for treatment [47-48]. During this work and via re-examining HIV-1 protease inhibitors and different anti bronchitis antibiotics we tried to find more effective inhibitors for COVID-19 treatment. Finally, we picked up the following order for inhibitory potency among anti-HIV-1 protease inhibitors:

Tipranavir>Indinavir>Atazanavir>Darunavir>Ritonavir>Amprenavir

Similarly, the following order for anti bronchitis was obtained

Cefditoren>Cefixime>Erythromycin>Clarithromycin

Considering this fact that all the drugs used in this study are approved by FDA for clinical application we hope that our findings considered for clinical trials in COVID-19 patients to overcome the world wide current outbreak of coronavirus.

## Acknowledgements

The author would like to express his thanks to the vice chancellor of research and technology of Shahid Chamran University of Ahvaz for providing financial support of this study under Research Grant No: SCU.SB98.477.